\documentclass{elsart}
\usepackage{graphicx,natbib,amssymb}
\usepackage{setspace}
\makeatletter
\def\elsartstyle{%
	\def\normalsize{\@setfontsize\normalsize\@xiipt{14.5}}
	\def\small{\@setfontsize\small\@xipt{13.6}}
	\let\footnotesize=\small
	\def\large{\@setfontsize\large\@xivpt{18}}
	\def\Large{\@setfontsize\Large\@xviipt{22}}
	\skip\@mpfootins = 18\p@ \@plus 2\p@
	\normalsize
}
\makeatother

\def\url#1{{\ttfamily\def\/{/\discretionary{}{}{}}#1}}

\begin{document}

\begin{frontmatter}
\title{High Resolution \'{E}chelle Spectroscopy of Two High Proper Motion Stars:
HD 102870 and BD+20 3603}

\begin{center}
\large
S.\,O.\,Solakc{\i}$^{\mathrm{1}}$,
T.\,\c{S}ahin$^{\mathrm{2}*}$,
C.\,Flynn$^{\mathrm{3}}$,
A.\,Dervi\c{s}o\u{g}lu$^{\mathrm{4}}$\\[12pt]

\small\itshape

$^1$Akdeniz University, Faculty of Science, Physics Department, 07058, Antalya, Turkey\\
$^2$Akdeniz University, Faculty of Science, Space Science and Technologies Department, 07058, Antalya, Turkey\\
$^3$Center for Astophysics amd Supercomputing, Swinburne University of Technology, Hawthorn, Australia\\
$^4$Department of Astronomy and Space Sciences, Erciyes University, 38039, Kayseri, Turkey\\

\end{center}

\thanks[email]{E-mail: timursahin@akdeniz.edu.tr; oktaysolakci@gmail.com}

\begin{abstract}

\noindent A chemical abundance analysis is made of two F type high proper
motion stars selected from the {\sc ELODIE} library. We use high resolution
(R=42\,000) and high signal to noise ratio (S/N=103, 36 $-$per$-$pixel)
\'{e}chelle spectra from the {\sc ELODIE} library of HD\,102870 and
BD\,+20\,3603, as two representative F type high proper motion stars, to
determine fundamental parameters and photospheric abundances of 16 chemical
elements including slow (s)$-$ and rapid (r)$-$process elements from Y to
Ba. The chemical composition and kinematic parameters of the stars imply that
they belong to different Galactic populations: we report HD\,102870, an IAU
standard radial velocity star, to be a thin disk star and BD\,+20\,3603, a
metal-poor HPM star, to be a halo star.

\end{abstract}

\end{frontmatter}

\section{Introduction}
\label{intro}

The study of unevolved late-type stars belonging to different Galactic
populations (the halo, the thick and thin disks) with large proper motions is
fundamental for our knowledge of the Galaxy's stellar content as well as for
reconstructing its chemical evolution. Late-type, metal-poor stars (the F to K
type dwarfs such as BD\,+20\,3603) with narrow and less blended spectral lines
in their spectra are excellent probes at medium resolution of the chemical
history of the earliest stellar populations. Large ongoing surveys of such
stars, such as HERMES/GALAH, ESO-Gaia and RAVE depend on accurately calibrated 
stellar parameters from such stars. 

We are currently undertaking a program of determining abundances for a large
number of elements from High Proper Motion (here after HPM) stars and
correlating these with the stellar Galactic orbital parameters. The project is
based on high resolution \'{e}chelle spectra obtained with the 1.9m
Observatoire de Haute-Provence telescope and spectrograph. Our whole HPM (-362
to +689 mas.yr$^{\rm -1}$ in RA; -899 to +339 mas.yr$^{\rm -1}$ in
Dec.)\footnote{Proper motions and distances (computed from parallaxes) are from
  {\sc SIMBAD} database at http://{\sc SIMBAD}.u-strasbg.fr/{\sc
    SIMBAD}/. Atmospheric parameters are from the {\sc ELODIE} archive at
  http://atlas.obs-hp.fr/{\sc ELODIE}/; Moultaka et al. (2004).} sample consists of 54 F-type stars,
along the {\sc ELODIE} library (Prugniel \& Soubiran 2001) with effective temperatures $4900 <$ T$_{\rm
  eff}$ $< 6900$ K, surface gravities $1.70 <$ log g $< 5.00$, metallicities
$-2.40 <$ [Fe/H] $< 1.00$, and distances $9 <$ d $< 283$ pc.  In this paper, we
aim to determine chemical abundances of two interesting program stars: HD
102870, a common star among IAU standard radial velocity stars as listed in the
Astronomical Almanac, also a {\sc GAIA} FGK Benchmark Star (GBS; Jofr\'{e} et al. 2014) and BD\,+20\,3603, a metal-poor HPM star. In Section 2,
we briefly give information on data selection process while Section 3 describes
the observations and the methods used for chemical abundance analysis. In
Section 4, we present results on their kinematics.  In Section 5, we report the
results of LTE based determination of model atmosphere parameters and elemental
abundances suggesting a thin disk and halo membership for HD\,102870 and
BD\,+20\,3603, respectively.

\section{Data Selection}

The {\sc ELODIE} (Soubiran 2003) library contains 1953 spectra for 1388 stars
with F and G types being the most common: the library contains 615 F type and
545 G type stars. Our first selection criterion for the program stars was
their spectral types. Among the F spectral type stars, we chose only the stars
with an HPM designation in the library, and removed the spectroscopic binaries
from the sample. Binarity is checked by using {\sc SIMBAD} database\footnote{It
  should be noted that comments on binarity is not reported in the {\sc ELODIE}
  archive.}. Thus, in total, we are left with 54 \'{e}chelle spectra for
spectroscopic analysis from the library.

 \begin{figure*}[!h]
 \centering \includegraphics*[width=14cm,height=10cm,angle=0]{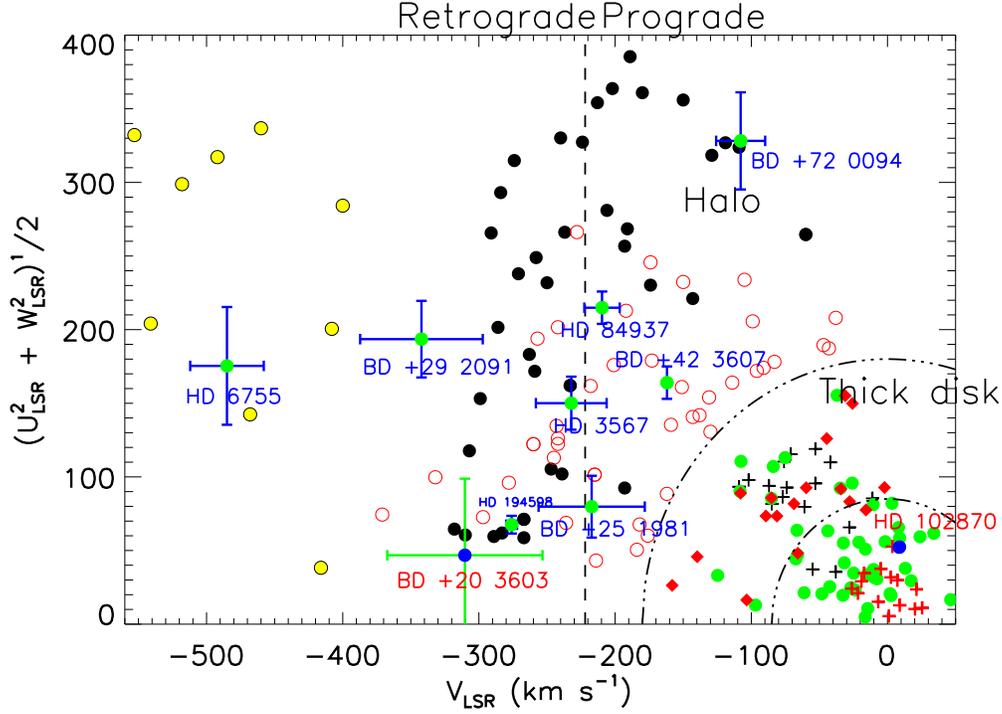}
 \caption{The Toomre energy diagram for program stars from (Nissen \& Schuster
   2010) and (Bensby et al. 2005). The positions of HD\,102870 and
   BD\,+20\,3603 are indicated.}
 \label{toomre}
 \end{figure*}

\section{Observation and Analysis}

High resolution (R=42\,000) and high signal-to-noise ratio (S/N$=150$) {\sc
  ELODIE} data provides a spectral coverage from 3900 to 6800 \AA\,. The
spectra were continuum normalized, wavelength calibrated and radial velocity
corrected by the data-reduction pipeline run at the telescope.  Prior to the
abundances analysis, the {\sc ELODIE} spectra were re-normalized, using a
in-house developed interactive normalization code {\sc INSS}\footnote{{\sc
    INSS}-Interactive Normalization of Stellar Spectra, developed under the
  {\sc TUBITAK} project (No:111T219) on "Radial Velocity measurements of RV
  Tauri-like IRAS Stars".} written in {\sc IDL}.

We have used {\sc LIME} - Line Measurements from \'{e}chelle Spectra (\c{S}ahin 2013) for
the line identification process. The current version of {\sc LIME} (ver. 3.0.) provides not
only the most probable identifications for the line of interest but also lists the related
atomic data (e.g. Rowland Multiplet Number-RMT, log$(gf)$, and Lower Level Excitation
Potential-LEP) that are compiled (e.g. from {\sc NIST} database) and fed into the code by
the user. Equivalent widths (EWs) are obtained using {\sc SPECTRE} (Sneden 1973). Finally,
we determine the chemical composition of the stars by using an {\sc LTE} line analysis code {\sc
MOOG} (Sneden 1973) and an {\sc ATLAS9} computed (ODFNEW) model atmospheres.  In the limit
that a line selection contains relatively weak lines (i.e. $12 \le EW \le 170$ m\AA\,), the
effective temperatures (T$_{\rm eff}$) is found by imposing the condition that the derived
abundance be independent of the lower level excitation potential (LEP). In the limit that
all lines have the same LEP and a similar wavelength, the microturbulence ($\xi$) is found
by requiring that the derived abundance be independent of reduced equivalent width (EW). For
abundance determinations of most of the HPM stars including HD\,102870 and
BD\,+20\,3603 in our {\sc ELODIE} sample, this procedure
was followed. For our sample of Fe\,{\sc I} lines, these two conditions are imposed
simultaneously since these atmospheric parameters are interdependent, an iterative procedure
is necessary. We determined the surface gravity ($log\,g$) by requiring ionization
equilibrium, e.g. that Fe\,{\sc I} and Fe\,{\sc II} lines produce the same iron abundance.
\footnote{The reason for this choice of Fe lines is that Fe lines are often quite numerous, even in
very metal-poor stars}. The details of abundance analysis and source of atomic data are the
same as in {\c S}ahin \& Lambert (2009) and {\c S}ahin et al. (2011).

\begin{figure*}[!ht]
\centering \includegraphics*[width=10cm,height=14cm,angle=270]{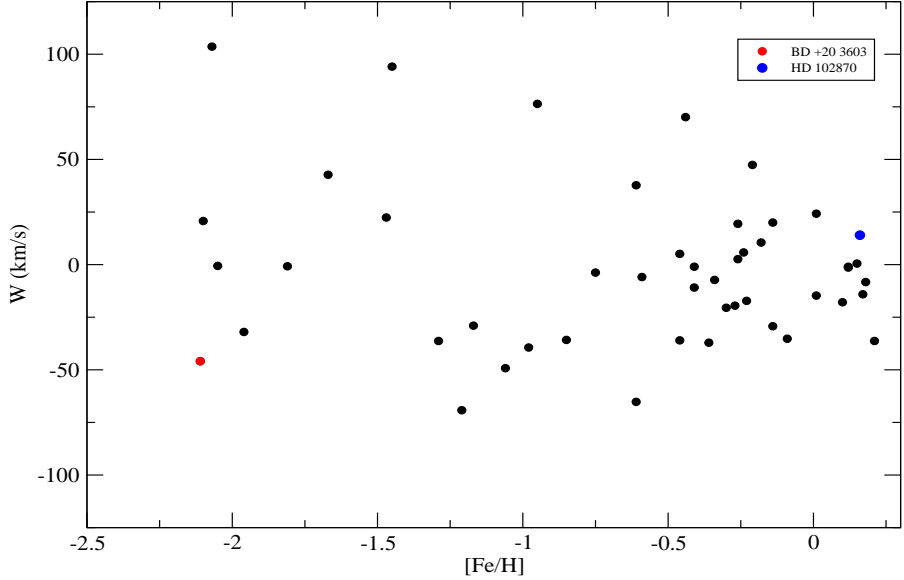}
\caption{Distribution of the 54 F-type HPM stars from the ELODIE archive in the
  plane metallicity - $W$ velocity.}
\label{feh_w}
\end{figure*}

\section{Space Velocities and Orbits}

In order to analyse the content of our HPM sample in terms of stellar
population, we compute the space velocities of the stars. Information on their
positions (from {\sc SIMBAD}), parallaxes and proper motions are from the new
reduction of the Hipparcos data (Van Leeuwen 2007). Radial velocities have been
computed by cross-correlation methodology (with an accuracy better than 0.1 km
s$^{-1}$). We calculate the Galactic space velocities U (toward Galactic
Center, l = 0$^{\circ}$), V (along the Galactic Rotation: l = 90$^{\circ}$, b =
0$^{\circ}$), W (towards the North Galactic Pole, b=90$^{\circ}$) and their
errors with respect to the local standard of rest using the method presented in
Johnson \& Soderblom (1987). Corrections for the solar motion (10.0, 5.25,
7.17) km s$^{-1}$ by Dehnen \& Binney (1998) were adopted.  Finally, using
those computed Galactic space velocities, a Toomre Energy Diagram consisting of
a complete list of HPM stars from the archive with a sample of stars from
Nissen \& Schuster (2010) (outer halo stars with black filled circle, inner
halo star with open red circles, and thick disk stars with crosses) and
Schuster (1993) (high velocity stars with yellow filled circle) was
compiled.(see Fig.1) Additional samples of thin (red plus sign), thick (red
filled diamond) disk stars from Bensby et al. (2005) with our sample of HPM
stars\footnote{The error bars (in blue) are indicated only for the stars with
  large velocity errors ($> 5$ km/s) in the total space velocity.} (filled
green circle) are also included. The dashed line in Figure 1. indicates zero
rotation in the Galaxy and the arcs correspond to V$_{\rm total}=$180 km/s and
87 km/s, the definitions of ``thick disk'' and ``disk'' (Nissen 2004). Figure
2. shows the distribution of the sample stars in the plane metallicity - $W$
velocity.

\noindent In order to complement the chemical abundances obtained in this
study, we also calculate Galactic orbital parameters for HD\,102870 and
BD\,+20\,3603 with the standard gravitational potentials described in the
literature (Miyamoto \& Nagai 1975; Hernquist 1990; Johnston et al.  1995;
Dinescu et al. 1999). The Solar Galactocentric distance and circular velocity
are taken to be 8.0 kpc and 220 km s$^{-1}$, respectively. For the orbital
integration, a fourth order symplectic integration technique by Yoshida (1990)
was applied. The typical orbital integration corresponds to 3 Gyr, and is
sufficient to evaluate the orbital elements of the HPM sample. Testing the code
for orbital parameters of 160 stars from Takeda et al. (2007) gives a mean
error of 0.01 in eccentricity and of 0.11 kpc in Galactocentric distance.

\begin{figure}[!ht]
\centering \includegraphics*[width=12cm,height=14cm,angle=270]{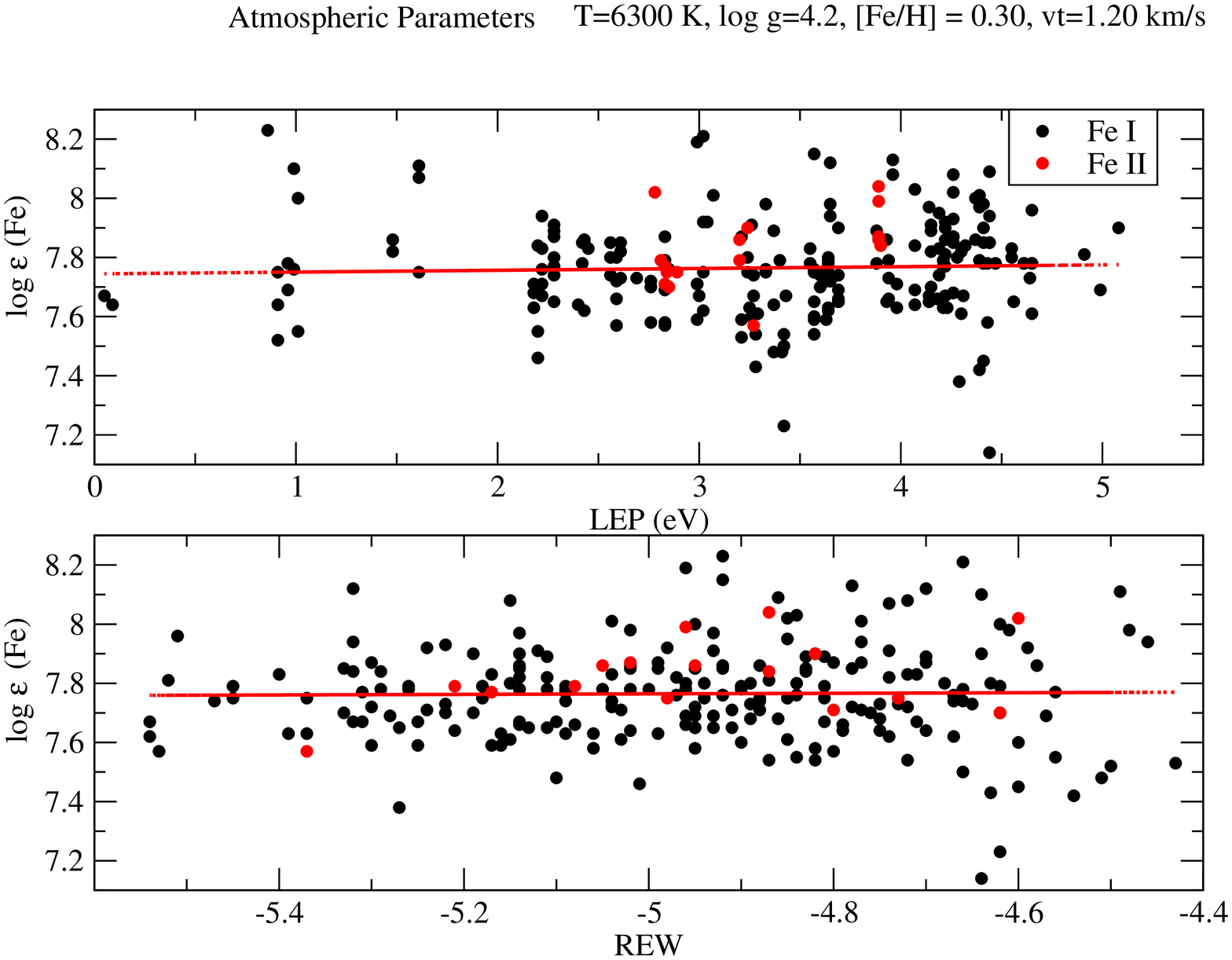}
\caption{An example for the determination of atmospheric parameters T$_{\rm
    eff}$ and $\xi$ using abundance (log$\epsilon$) as a function of both lower
  level excitation potential (LEP, upper panel) and reduced EW (REW; log
  (EW/$\lambda$), lower panel). The red solid line in the all panels is the
  least-square fit to the data. The zero slope of this line is for T$_{\rm
    eff}$= 6300 K.}
\label{m_param_}
\end{figure}

\section{Results and conclusions}

\subsection{HD102870: a radial velocity standard star}

\noindent Chemical abundances based on our equivalent width analysis of {\sc
  IAU} standard radial velocity star (as listed in the Astronomical Almanac):
HD\,102870 for the following model parameters: T$_{\rm eff}$ = $6300\pm150$ K,
log\,g = $4.2\pm0.3$, [Fe/H] = $0.30\pm0.16$, and $\xi$ = $1.2\pm0.5$ km
s$^{-1}$ are obtained. Figure 3 presents an example plot for the determination
of atmospheric parameters. The spectrum of HD\,102870 is presented in Figure
4. Intensity differences between {\sc ELODIE} and McDonald spectra in neutral
and ionized lines of Si, Ca, Sc, Cr, Mn, Fe, and Co are interesting to
note. Figure 5 shows a comparison of model parameters obtained in this study to
those listed in the literature.

The final elemental abundances log $\epsilon$(X) averaged over the sets of
measured lines are listed in Table 1, where the first column gives species,
second gives logarithmic elemental abundances and third gives element over iron
ratios for the star\footnote{[X/Fe]=[log$\epsilon$(X)-log$\epsilon$(Fe)]$_{\rm
    star}$ - [log$\epsilon$(X)-log$\epsilon$(Fe)]$_{\rm \odot}$}. Column four
and five give element over iron ratios for thin disk stars from Bensby et
al. (2005) and Solar abundances from Asplund et al. (2009). The number of lines
used in the analysis are presented in the column six.

\begin{figure}[!ht]
\centering
\includegraphics*[width=14cm,height=10cm,angle=0]{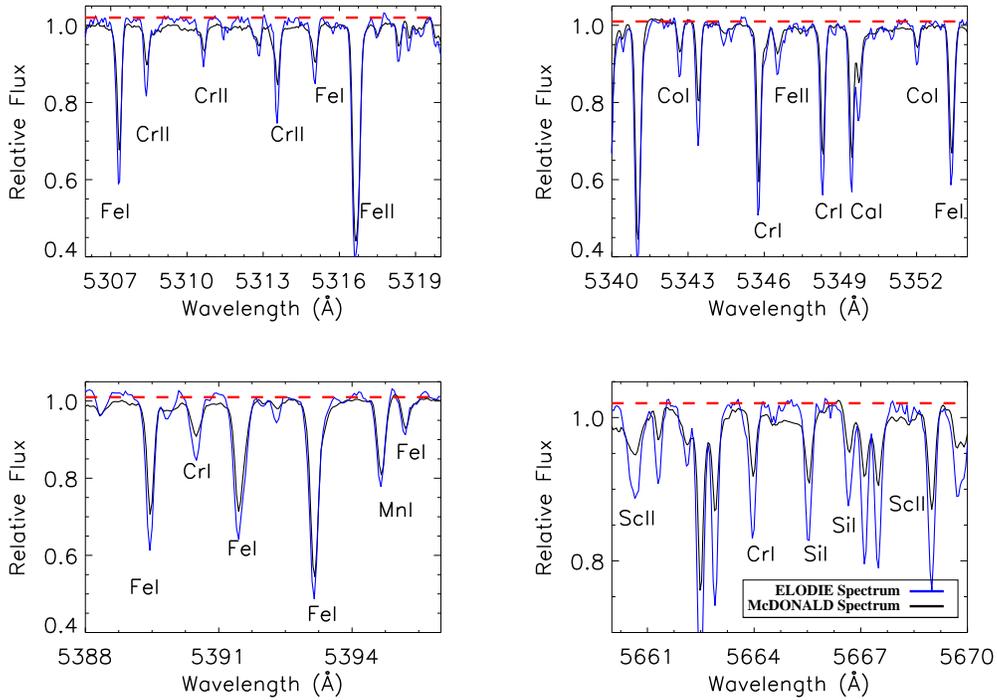}
\caption{An example spectrum of HD\,102870: {\sc ELODIE} spectrum (in blue) and
  McDonald (in black; obtained in 2009, April 13 with 2.1m Otto Struve
  Telescope).}
\label{e_spec_HD102870}
\end{figure}

On purely chemical grounds, the metallicity of HD\,102870 implies a thin disk
membership for the star ($-0.2 >$ [Fe/H] for the thin disk, Lambert
1988). The $\alpha$ element abundance for the star with [$\alpha$/Fe]$\approx$-0.1 is also typical for a thin disk star (Navarro et al. 2011). Further confirmation for thin disk membership came from kinematics of
the star.

\begin{figure}[!ht]
\centering
\includegraphics*[width=14cm,height=10cm,angle=0]{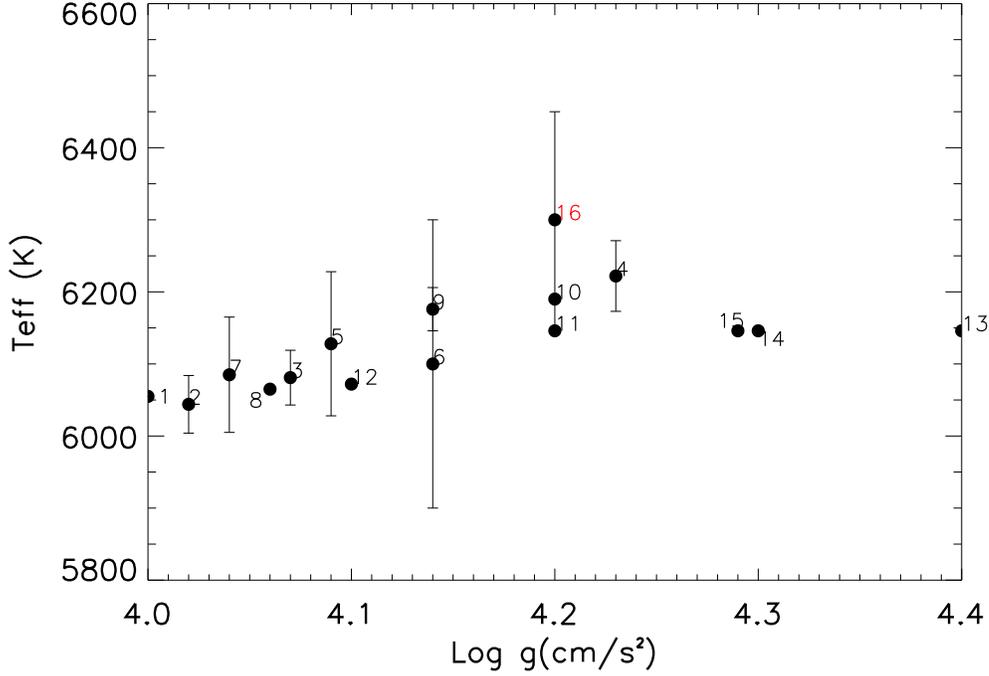}
\caption{Comparison of the model parameters from 1- Mishenina et al.(2013),2-
  Maldonado et al.(2012),3- Prugniel et al.(2011), 4- Ghezzi et al.(2010), 5-
  Takeda (2007), 6- Mallik et al.(1998), 7- Fuhrmann et al.(1998),
  8-Gratton et al.(1996), 9- Edvardsson et al.(1993), 10- Balachandran et
  al.(1990), 11-Thevenin et al.(1986), 12-Boesgard et al.(1986), 13-Edvardsson
  et al.(1984), 14-Gehren et al.(1981), 15-Baschek et al.(1967) for
  HD102870. Model parameters obtained in this study (16) is also indicated.}
\label{m_param_HD102870}
\end{figure}

\noindent The Galactic space velocities to be used to determine membership
status of the star has been calculated as $(U,V,W)$ = ($50.4\pm0.1$,
$8.5\pm0.1$, $14.0\pm0.1$) km/s. The Toomre energy diagram is shown in Figure
1. also verifies the thin disk membership for the star.

\begin{table}
\caption[]{Chemical abundances of HD\,102870.}
$$
\begin{array}{l|l|c|c|c|c|l}
           \hline
           \hline
    Elem. & \log\epsilon(X) &  [X/Fe] & [X/Fe]_{\rm B05} &  [X/Fe]_{\rm M04} &\log\epsilon_{\rm \odot} &  Note \\
    \hline
    C\,{\sc I}    &    8.61\pm0.12 &$-$0.12  &   --   &   --   &8.43 &   5   \\
    Na\,{\sc I}   &    6.51\pm0.05 &$-$0.03  &  0.06  &   --   &6.24 &   2   \\
    Si\,{\sc I}   &    7.73\pm0.15 &$-$0.08  &  0.05  &   --   &7.51 &   7   \\
    Si\,{\sc II}  &    7.81\pm0.09 &  0.00   &  0.05  &   --   &7.51 &   2   \\
    Ca\,{\sc I}   &    6.46\pm0.19 &$-$0.18  &  0.05  &$-$0.02 &6.34 &  17   \\
    Ca\,{\sc II}  &    6.40\pm0.00 &$-$0.24  &  0.05  &   --   &6.34 &   1   \\
    Sc\,{\sc II}  &    3.59\pm0.10 &  0.14   &  --    &   --   &3.15 &   7   \\
    Ti\,{\sc I}   &    5.18\pm0.12 &$-$0.07  &  0.02  &   --   &4.95 &  23   \\
    Ti\,{\sc II}  &    5.42\pm0.25 &  0.17   &  0.05  &   --   &4.95 &  12   \\
    V\,{\sc I}    &    4.42\pm0.29 &  0.19   &  --    &   --   &3.93 &   7   \\
    V\,{\sc II}   &    4.25\pm0.00 &  0.02   &  --    &   --   &3.93 &   1   \\
    Cr\,{\sc I}   &    5.97\pm0.20 &  0.03   &$-$0.01 &   --   &5.64 &  28   \\
    Cr\,{\sc II}  &    6.08\pm0.24 &  0.14   &$-$0.01 &   --   &5.64 &   6   \\
    Mn\,{\sc I}   &    5.73\pm0.20 &  0.00   &  --    &   --   &5.43 &  13   \\
    Fe\,{\sc I}   &    7.80\pm0.17 &  0.00   &  0.00  &    --  &7.50 & 216   \\
    Fe\,{\sc II}  &    7.83\pm0.12 &  0.00   &  0.00  &    --  &7.50 &  16   \\
    Ni\,{\sc I}   &    6.51\pm0.12 &  0.01   &$-$0.01 & 0.04   &6.22 &  48   \\
    Y\,{\sc II}   &    2.47\pm0.04 &$-$0.04  &  --    &$-$0.07 &2.21 &   2   \\
    Zr\,{\sc II}  &    2.84\pm0.00 &$-$0.04  &  --    &$-$0.04 &2.58 &   1   \\
    Ba\,{\sc II}  &    2.45\pm0.00 &$-$0.01  &  --    &$-$0.03 &2.18 &   1   \\
\hline				
\hline
\end{array}
$$
\end{table}

Finally, we compute thin and thick disk probabilities for the star on the basis
of its spatial velocity. Those respective thin and thick disk membership
probabilities as 0.95 (Pr1) and 0.05 (Pr2), with the recipe given by Mishenina
et al. (2004) and also using the membership test by Bensby et al. (2005),
clearly indicates to a thin disk membership with a thick-disk-to-thin-disk (TD / D)
probability of 0.02. We also computed the orbit of the star. The computed
apogalactic and perigalactic distances of the star are found to be
$9.63\pm0.01$ kpc and $7.11\pm0.01$ kpc and a maximum height excursion above
the Galactic plane of $230\pm10$ pc. An orbital eccentricity of 0.15 is
obtained. Additional test by isochrones of Bertelli (1994) for estimation of
age gives an age of $\approx$9 Gyrs for this thin disk star.\footnote{Age
  estimation is based on atmospheric parameters: T$_{\rm eff}$, log\,g and
  [Fe/H] and calculations are performed on the basis of scaled-solar evolution
  models (see Allende Prieto et al. 2004; Reddy et al. 2006, and Ramirez et
  al. 2007).}


\begin{figure}[!ht]
\centering \includegraphics*[width=12cm,height=14cm,angle=270]{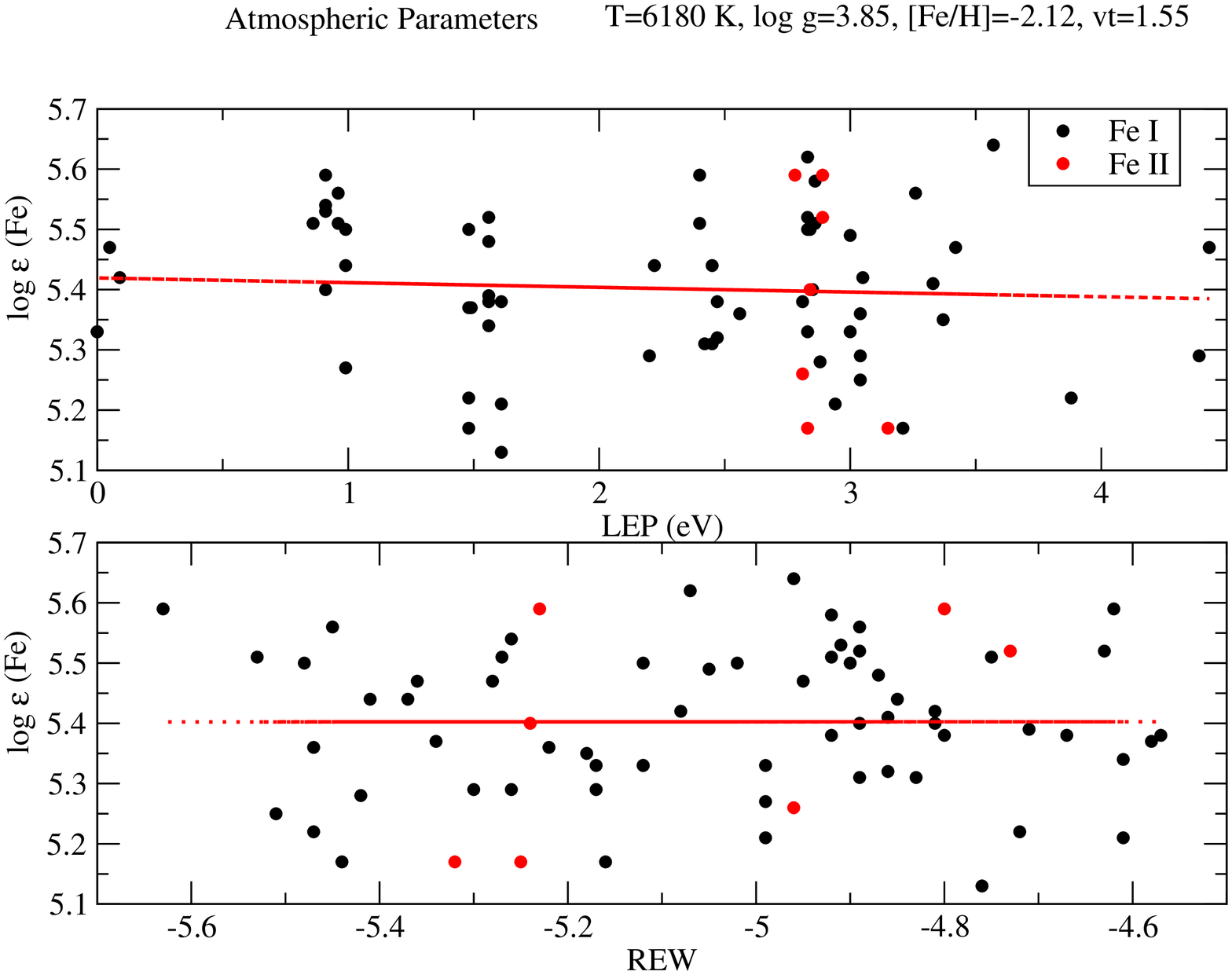}
   \caption{Determination of atmospheric parameters for BD\,+20\,3603. The red
     solid line in the all panels is the least-square fit to the data. The zero
     slope of this line is for T$_{\rm eff}$= 6180 K}
\label{f1_nuv_g}
\end{figure}

\subsection{BD +20 3603: a metal-poor high-proper motion star}

The chemical abundances and kinematic analysis of metal-poor high-proper motion
star, BD\,+20\,3603, was performed . In Table 2, we present results on
equivalent width analysis of the spectrum for the following model parameters:
T$_{\rm eff}$ = 6180$\pm$150 K, $\log\,g = 3.85\pm0.3$, [Fe/H]$ =
-2.12\pm0.12$, and $\xi = 1.55\pm0.5$.  Figure 6 presents an example plot for
the determination of atmospheric parameters.  An example spectrum of the star
is presented in Figure 7. Figure 8 presents a comparison of model parameters
obtained in this study to those listed in the literature. Table 2 presents the
elemental abundances with element over iron ratios from Bai et al. (2004) in
the fourth column.  To check our determination of abundances and compare line
to-line scatter values, we used equivalent widths, reported line list and
atomic data from a low resolution spectrum (R$\approx$15000) of the star by Bai
et al. (2004).

\begin{figure}[!ht]
\centering
\includegraphics*[width=14cm,height=13cm,angle=0]{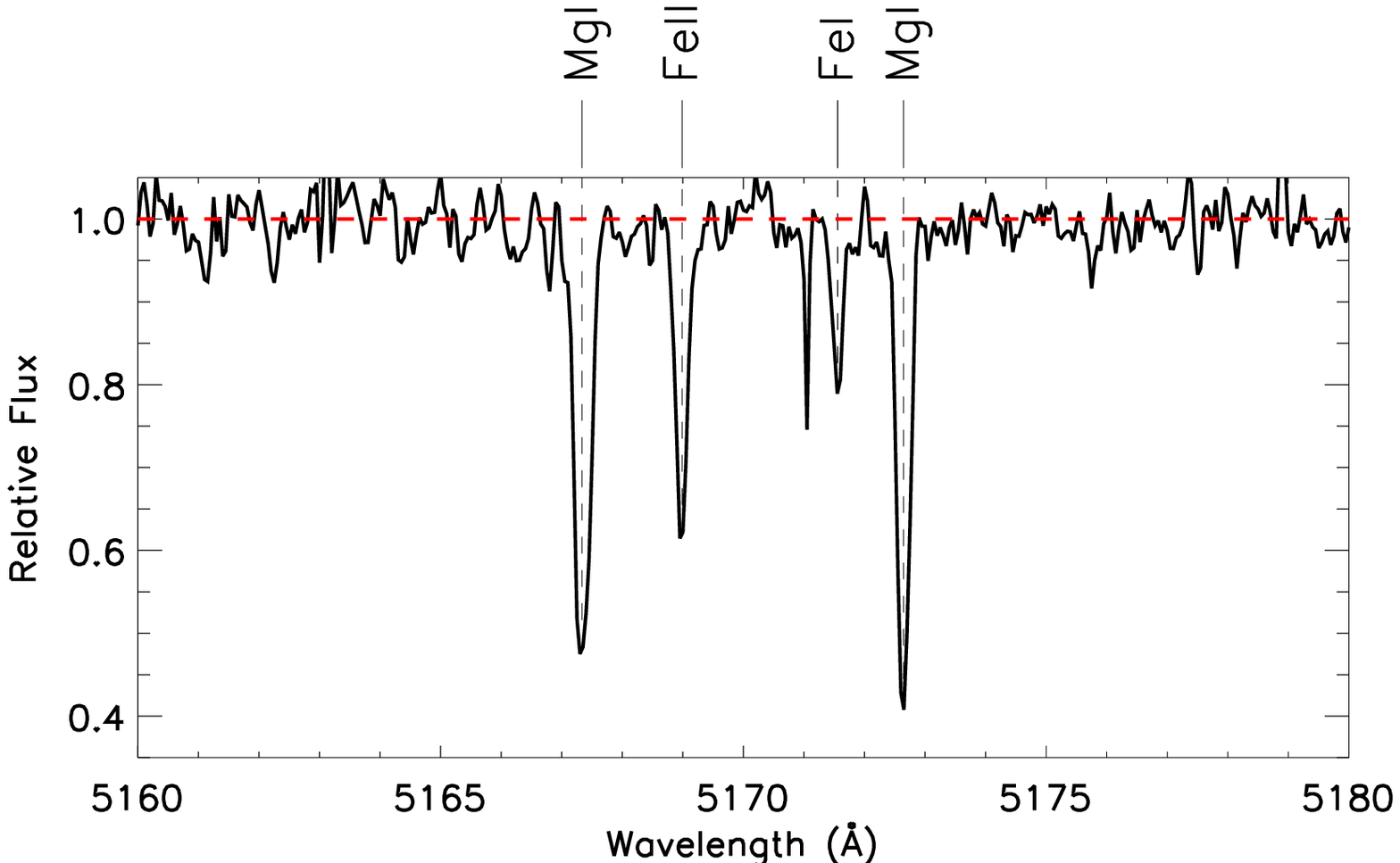}
   \caption{Example spectrum plot for BD\,+20\,3603 from the {\sc ELODIE}
     archive.}
\label{f1_nuv_g}
\end{figure}

\begin{figure}[!ht]
\centering
\includegraphics*[width=14cm,height=10cm,angle=0]{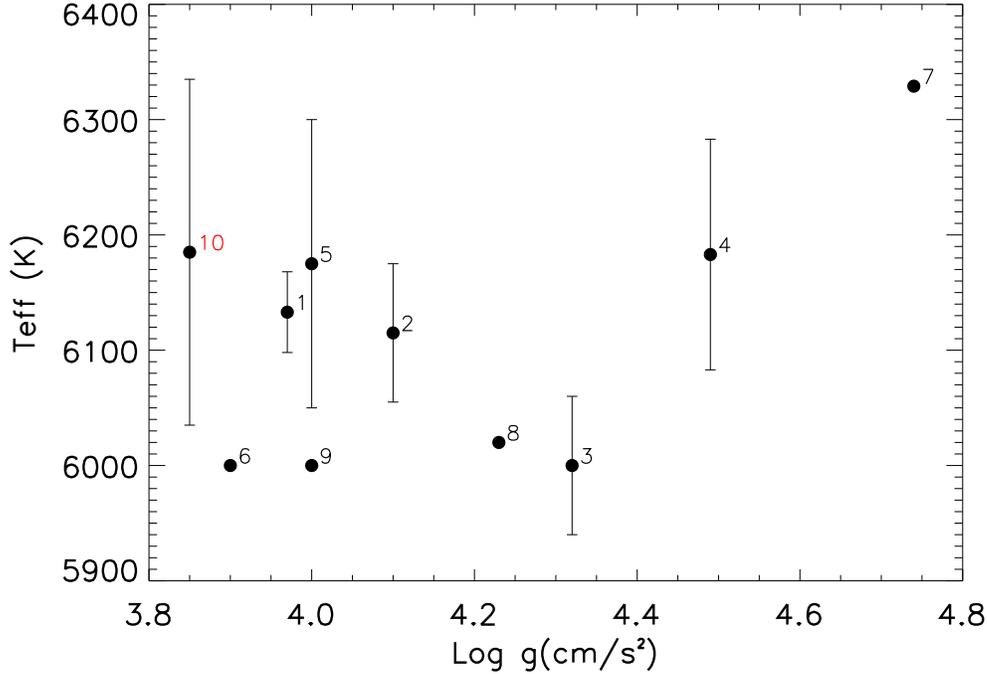}
  \caption{Comparison of the model parameters from 1- Wu et al.(2011), 2-
    Prugniel et al.(2011), 3- Cenarro et al.(2007), 4- Zhang \& Zhao (2005), 5-
    Fullbright et al.(2000), 6- Carney et al.(1997), 7-Gratton et al.(1996), 8-
    Tomkin et al.(1992), 9- Peterson (1981) for BD\,+20\,3603. Our result
    (10) is also indicated.}
\label{f1_nuv_g}
\end{figure}

\begin{table}
\caption[]{Chemical abundances of BD\,+20\,3603.}
$$
\begin{array}{l|l|c|c|c|l}
           \hline
           \hline
    Elem. & \log\epsilon(X) &  [X/Fe] & [X/Fe]_{\rm Bai} & \log\epsilon_{\rm \odot} &  Note \\
    \hline
    Mg\,{\sc I}   &    5.71\pm0.03 &  0.21 & 0.36 & 7.60 & 4     \\
    Ca\,{\sc I}   &    4.59\pm0.12 &  0.35 & 0.37 & 6.34 &  11   \\
    Sc\,{\sc II}  &    1.03\pm0.20 &$-$0.02& 0.17 & 3.15 &   4   \\
    Ti\,{\sc I}   &    3.32\pm0.08 &  0.47 & 0.53 & 4.95 & 2     \\ 
    Ti\,{\sc II}  &    3.18\pm0.09 &  0.35 & 0.53 & 4.95 & 11    \\ 
    Cr\,{\sc I}   &    3.72\pm0.08 &  0.18 & 0.05 & 5.64 &  3    \\
    Fe\,{\sc I}   &    5.40\pm0.12 &  0.00 & 0.00 & 7.50 & 63    \\
    Fe\,{\sc II}  &    5.39\pm0.19 &  0.00 & 0.00 & 7.50 &  7    \\
    Ni\,{\sc I}   &    4.33\pm0.00 &  0.21 & --   & 6.22 &  1    \\
    Sr\,{\sc II}  &    0.15\pm0.00 &$-$0.62& --   & 2.87 & 1     \\
\hline				
\hline
\end{array}
$$
\end{table}

Bai et al. (2004) do not report line-by-line abundances for the elements
studied for equivalent width analysis but only present the (mean) element over
iron ratios.  So, in order to perform this test for comparison purposes, a
model atmosphere using the following model parameters of Bai et al. (2004)
(T$_{\rm eff} = 6138$ K, $\log\,g = 3.95 $cm\,s$^{-2}$, [Fe/H] $= -2.09$, $\xi
= 1.50$ km\,s$^{-1}$ ) was computed. Neutral and ionized Fe abundances are
found as follows: [Fe\,{\sc I}] $= 5.30\pm0.20$ and [Fe\,{\sc II}]$ =
5.33\pm0.14$ with a line-to-line scatter as high as 0.5 dex. Using a {\sc
  KUROLD} instead of {\sc NEWODF} model increased neutral and ionized iron
abundances by only 0.09 dex and 0.07 dex, respectively. These values are in
accordance with our determination of iron abundance: [Fe\,{\sc I}] =
5.40$\pm$0.12, [Fe\,{\sc II}] $= 5.39\pm0.19$. The test is performed with a
$\approx$3 times higher resolution spectrum from the {\sc ELODIE} library with
our determination of equivalent widths provided as input and a similar source
of atomic data (i.e. {\sc NIST}) were used. The overabundances of magnesium ([Mg/Fe]=0.21) and
calcium ([Ca/Fe]=0.35) as tracers of $\alpha$-process elements with [$\alpha$/Fe]$\approx$0.3 are expected for a halo star
(Magain 1987, 1989; Gratton \& Sneden 1987, 1988, 1991; Nissen et al. 1994; Fuhrmann et al. 1995; McWilliam et al. 1995). The computed Galactic velocity components for BD\,+20\,3603 are found
to be $(U, V, W) = (-9\pm51, -311\pm57, -46\pm13)$ km\,s$^{-1}$. The method of
Bensby (2005) for BD\,+20\,3603 implies a halo membership for the star. The
apogalactic and perigalactic distances of 8.05$\pm$0.60 kpc and 2.16$\pm$2.0
kpc with a Zmax of 890$\pm$600 pc and an orbital eccentricity of 0.58 indicates
to a halo membership for the star. The isochrones by Bertelli (1994)
corraborates the membership status of the star as a halo star with an age of
$\approx$9 Gyr. The star, based on elemental abundances, metallicity (based on
metallicity criteria for halo objects as [Fe/H] $< -1.0$ by Lambert 1988) and
its kinematics supported with orbital computations seem to be a halo star.

\section{Future Work}

Focusing on chemistry of metal-poor and radial velocity standard stars with
relatively high space motions with a help of their kinematics will not only
extent our understanding of earliest type of stellar populations but also
improve accuracy of radial velocity measurements. Our report of relatively
higher metallicity of HD\,102870 as an IAU standard radial velocity star is a
unique example. Intensity differences between {\sc ELODIE} and {\sc McDonald}
spectra of the star in neutral and ionized Si, Ca, Sc, Cr, Mn, Fe, and Co lines
are interesting to note and needs to be investigated further by increasing the
number of HPM stars for spectroscopic analysis. This also implys an important
fact that special care has to be taken when using the radial velocity standard
stars with HPM designation in studies regarding precise measurements of radial
velocities. So the question inspired us to pursue this research remains
partially unanswered. The analysis of whole sample continues.  For some of the
program star with relatively low signal-to-noise ratio, additional spectra are
required. A dedicated web page containing results on abundances and kinematical
properties for whole HPM sample is also being prepared. We plan to extend our
study to G type HPM stars from the {\sc ELODIE} library in order to convey
effects of HPM nature of a star on its chemistry in more detail and to reveal possible correlation between kinematics and
abundance ratios.

\section{Acknowledgment}

TS  acknowledge the financial support by TÜBITAK (TBAG-1001; Project No: 111T219).

\end{document}